\documentclass[%
 reprint,
superscriptaddress,
]{revtex4-1}

\usepackage{mathrsfs}
\usepackage{graphicx}
\usepackage{dcolumn}
\usepackage{bm}
\usepackage{amsmath,amssymb, esint}
\usepackage{subcaption}
\usepackage{comment}
\usepackage{bm,wasysym,color}
\usepackage[version=4]{mhchem} 
\usepackage{textgreek}
\usepackage[ labelfont = {footnotesize, bf, sf}, figurename=Figure, format=plain, justification=raggedright]{caption}
\usepackage{tikz}
\usepackage{comment}
\usepackage{pgf,pgfplots}
\pgfplotsset{compat=1.3}
\usetikzlibrary{intersections}

\begin{document}
\pgfkeys{%
	  	/pgf/number format/set thousands separator = {,}}


\title{\sffamily\Large Nanophotonic Pockels modulators on a silicon nitride platform}
\affiliation{
Photonics Research Group, INTEC Department, Ghent University-imec, Ghent B-9000, Belgium
}
\affiliation{
Liquid Crystals and Photonics Group, ELIS Department, Ghent University, Ghent B-9000, Belgium
}
\affiliation{
IDLab, INTEC Department, Ghent University-imec, Ghent B-9000, Belgium
}
\affiliation{
Center for Nano- and Biophotonics (NB-Photonics), Ghent University, Ghent B-9000, Belgium
}
\author{\sffamily Koen Alexander}
\thanks{\textbf{These authors contributed equally to this work}}
\affiliation{
Photonics Research Group, INTEC Department, Ghent University-imec, Ghent B-9000, Belgium
}
\affiliation{
Center for Nano- and Biophotonics (NB-Photonics), Ghent University, Ghent B-9000, Belgium
}
\author{\sffamily John P. George}
\thanks{\textbf{These authors contributed equally to this work}}
\affiliation{
Photonics Research Group, INTEC Department, Ghent University-imec, Ghent B-9000, Belgium
}
\affiliation{
Liquid Crystals and Photonics Group, ELIS Department, Ghent University, Ghent B-9000, Belgium
}
\affiliation{
Center for Nano- and Biophotonics (NB-Photonics), Ghent University, Ghent B-9000, Belgium
}
\author{\sffamily Jochem Verbist}
\affiliation{
Photonics Research Group, INTEC Department, Ghent University-imec, Ghent B-9000, Belgium
}
\affiliation{
IDLab, INTEC Department, Ghent University-imec, Ghent B-9000, Belgium
}
\affiliation{
Center for Nano- and Biophotonics (NB-Photonics), Ghent University, Ghent B-9000, Belgium
}
\author{\sffamily Kristiaan Neyts}
\affiliation{
Liquid Crystals and Photonics Group, ELIS Department, Ghent University, Ghent B-9000, Belgium
}
\affiliation{
Center for Nano- and Biophotonics (NB-Photonics), Ghent University, Ghent B-9000, Belgium
}
\author{\sffamily Bart Kuyken}
\author{\sffamily Dries Van Thourhout}\email{Dries.VanThourhout@UGent.be}
\affiliation{
Photonics Research Group, INTEC Department, Ghent University-imec, Ghent B-9000, Belgium
}
\affiliation{
Center for Nano- and Biophotonics (NB-Photonics), Ghent University, Ghent B-9000, Belgium
}
\author{\sffamily Jeroen Beeckman}\email{Jeroen.Beeckman@UGent.be}
\affiliation{
Liquid Crystals and Photonics Group, ELIS Department, Ghent University, Ghent B-9000, Belgium
}
\affiliation{
Center for Nano- and Biophotonics (NB-Photonics), Ghent University, Ghent B-9000, Belgium
}

\date{\today}

\pacs{42,78}
\begin{abstract}
Silicon nitride (SiN) is emerging as a competitive platform for CMOS-compatible integrated photonics. However, active devices such as modulators are scarce and still lack in performance. Ideally, such a modulator should have a high bandwidth, good modulation efficiency, low loss, and cover a wide wavelength range. Here, we demonstrate the first electro-optic modulators based on ferroelectric lead zirconate titanate (PZT) films on SiN, in both the O- and the C-band. Bias-free operation, bandwidths beyond 33 GHz and data rates of 40 Gbps are shown, as well as low propagation losses ($\alpha\approx 1$ dB/cm).  A $V_\pi L\approx$ 3.2 Vcm is measured. Simulations indicate that values below 2 Vcm are achievable. This approach offers a much-anticipated route towards high-performance phase modulators on SiN.
\end{abstract}                            
                              
\maketitle
\begin{figure*}[!ht]
\centering
	\includegraphics{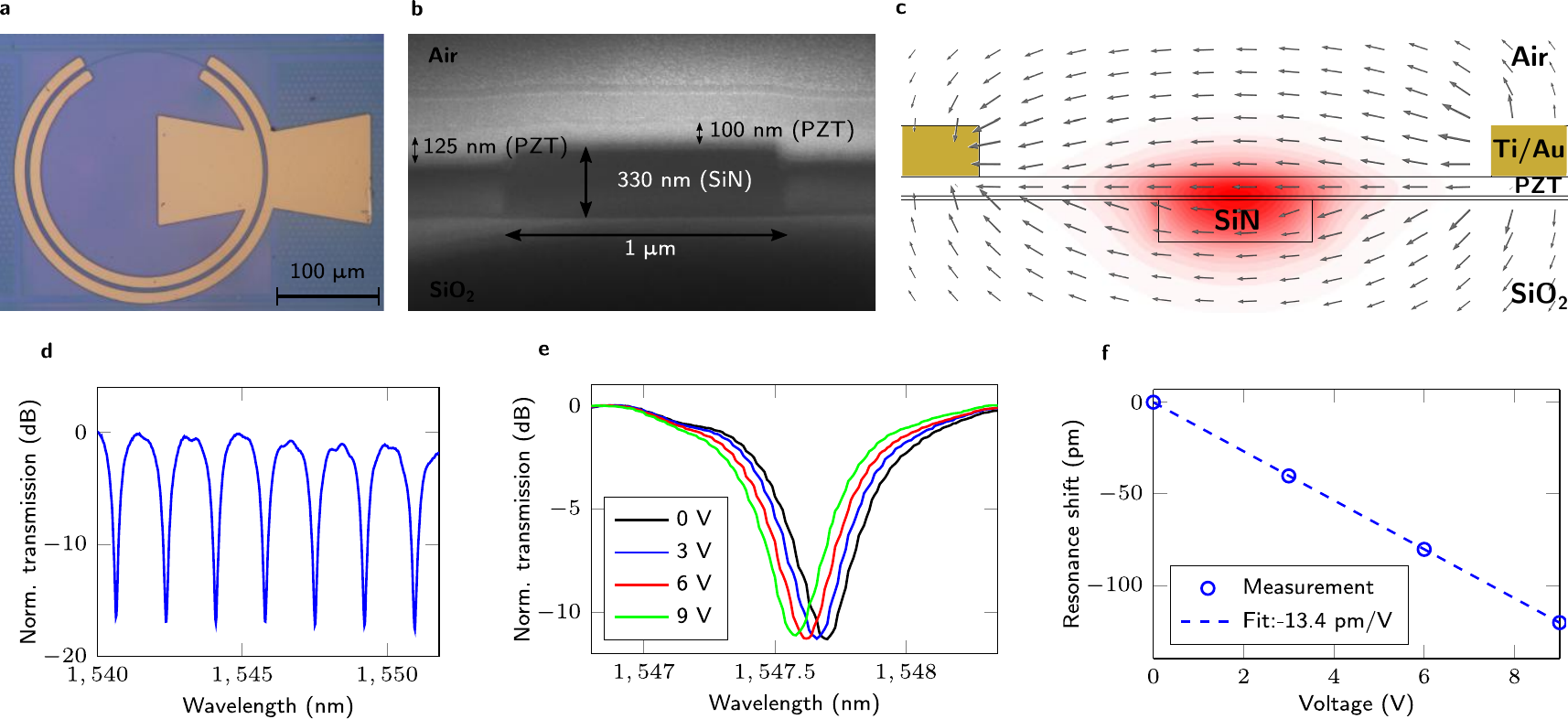}
  \caption{\sffamily\footnotesize \textbf{Design and static response of a C-band ring modulator.} \textbf{a}, Top view of a PZT-on-SiN ring modulator. \textbf{b}, Cross-section of a PZT-covered SiN waveguide. \textbf{c}, Schematic of the PZT-covered \ce{SiN} waveguide, the fundamental TE optical mode at is plotted in red. The quiver plot shows the applied electric field distribution between the electrodes. PZT thickness, waveguide width and gap between the electrodes are respectively 150 nm, 1200 nm and 4 \textmugreek m. \textbf{d}, Normalized transmission spectrum of a C-band ring modulator. \textbf{e}, Transmission spectra for different DC voltages. \textbf{f}, Resonance wavelength shift versus voltage applied across the PZT, including a linear fit. }\label{fig:fig1}
\end{figure*}

The exponential increase in data traffic requires high capacity optical links. A fast, compact, energy efficient, broadband optical modulator is a vital part of such a system. Modulators integrated with silicon (Si) or silicon nitride (SiN) platforms are especially promising, as they leverage CMOS fabrication techniques. This enables high yield, low-cost and scalable photonics, and a route towards co-integration with electronics \cite{Sun2015single}. SiN-based integrated platforms offer some added advantages compared to silicon-on-insulator (SOI), such as a broader transparency range \cite{rahim2017expanding}, a lower propagation loss \cite{levy2010cmos,bauters2011ultra}, significantly lower nonlinear losses \cite{rahim2017expanding, moss2013new}, and a much smaller thermo-optic coefficient \cite{rahim2017expanding}. Therefore, phase modulators on SiN in particular would open new doors in other fields as well, such as nonlinear  and quantum optics \cite{moss2013new, ramelow2015silicon, kahl2015waveguide}, microwave photonics \cite{zhuang2015programmable}, optical phased arrays for LIDAR or free-space communication \cite{poulton2017large}, and more.

State-of-the-art silicon modulators rely on phase modulation through free carrier plasma dispersion in p-n \cite{reed2014high}, p-i-n \cite{xu2005micrometre} and MOS \cite{liu2004high} junctions. Despite being relatively fast and efficient, these devices suffer from spurious amplitude modulation and high insertion losses. Alternative approaches are based on heterogeneous integration with materials such as III-V semiconductors \cite{hiraki2017heterogeneously, han2017efficient}, graphene \cite{liu2011graphene, sorianello2018graphene}, electro-optic organic layers \cite{alloatti2014100}, germanium \cite{srinivasan201656} or epitaxial \ce{BaTiO3} (BTO) \cite{abel2013strong, xiong2014active, Eltes2017a}.

Most of these solutions are not viable using SiN. Due to its insulating nature, plasma dispersion effects and many approaches based on co-integration with III-V semiconductors, graphene, and organics, which rely on the conductivity of doped silicon waveguides, cannot be used. The inherent nature of deposited SiN further excludes solutions using epitaxial integration. Finally, SiN is centrosymmetric, hampering Pockels-based modulation in the waveguide core itself, in contrast to  aluminum nitride \cite{xiong2012low}, or lithium niobate \cite{wang2018nanophotonic}.  Nonetheless, modulators on SiN exist. Using double-layer graphene, Phare et al. achieved high speed electro-absorption modulation \cite{phare2015graphene} and using piezoelectric PZT thin films, phase modulators based on stress-optic effects \cite{hosseini2015stress} and geometric deformation \cite{Jin2018piezoelectrically} have been demonstrated, albeit with sub-MHz electrical bandwidth.

In this work, we use a novel approach for co-integration of thin-film PZT on SiN \cite{george2015lanthanide}. An intermediate low-loss lanthanide-based layer is used as a seed for the PZT deposition, as opposed to the highly absorbing Pt-based seed layers used conventionally \cite{hosseini2015stress, Jin2018piezoelectrically}, enabling direct deposition of the layer on top of SiN waveguides.

We demonstrate the first efficient high speed phase modulators on a SiN platform, with bias-free operation, modulation bandwidths exceeding 33 GHz in both the O- and C-band, and data rates up to 40 Gbps. We measure propagation losses down to 1 dB/cm and half-wave voltage-length products $V_\pi L$ down to 3.2 Vcm for the PZT-on-SiN waveguides. Moreover, based on simulations we argue that the $V_\pi L$ can be considerably reduced by optimizing the waveguide cross-section, without significantly increasing the propagation loss. Pure phase modulation also enables complex encoding schemes (such as QPSK), which are not easily achievable with absorption modulation. These results not only strongly improve upon what is currently possible in SiN, they are also on par with the state-of-the-art modulators in silicon-on-insulator \cite{reed2014recent}.
\begin{figure}[h]
\centering
  	\includegraphics{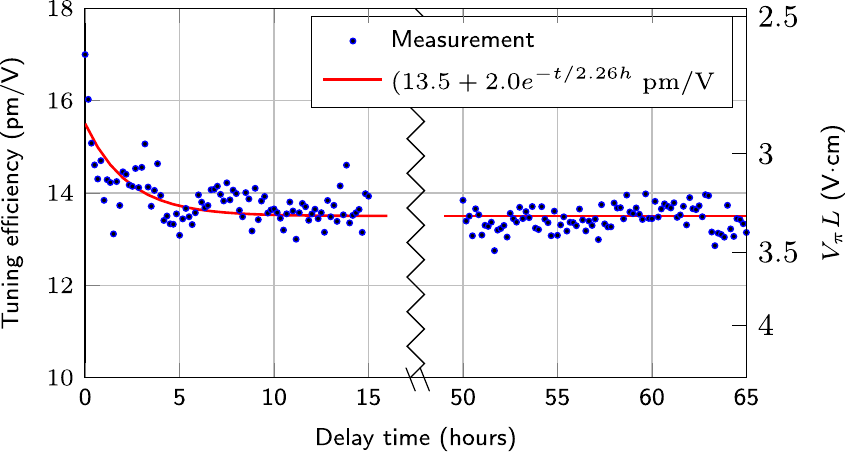}
  \caption{\sffamily\footnotesize \textbf{Poling stability of the electro-optic film.} Tuning efficiency (C-band ring) as a function of time after poling. The axis on the right shows the estimated corresponding $V_\pi L$.}\label{fig:fig2}
\end{figure}
\begin{figure*}[!ht]
\centering
	\includegraphics{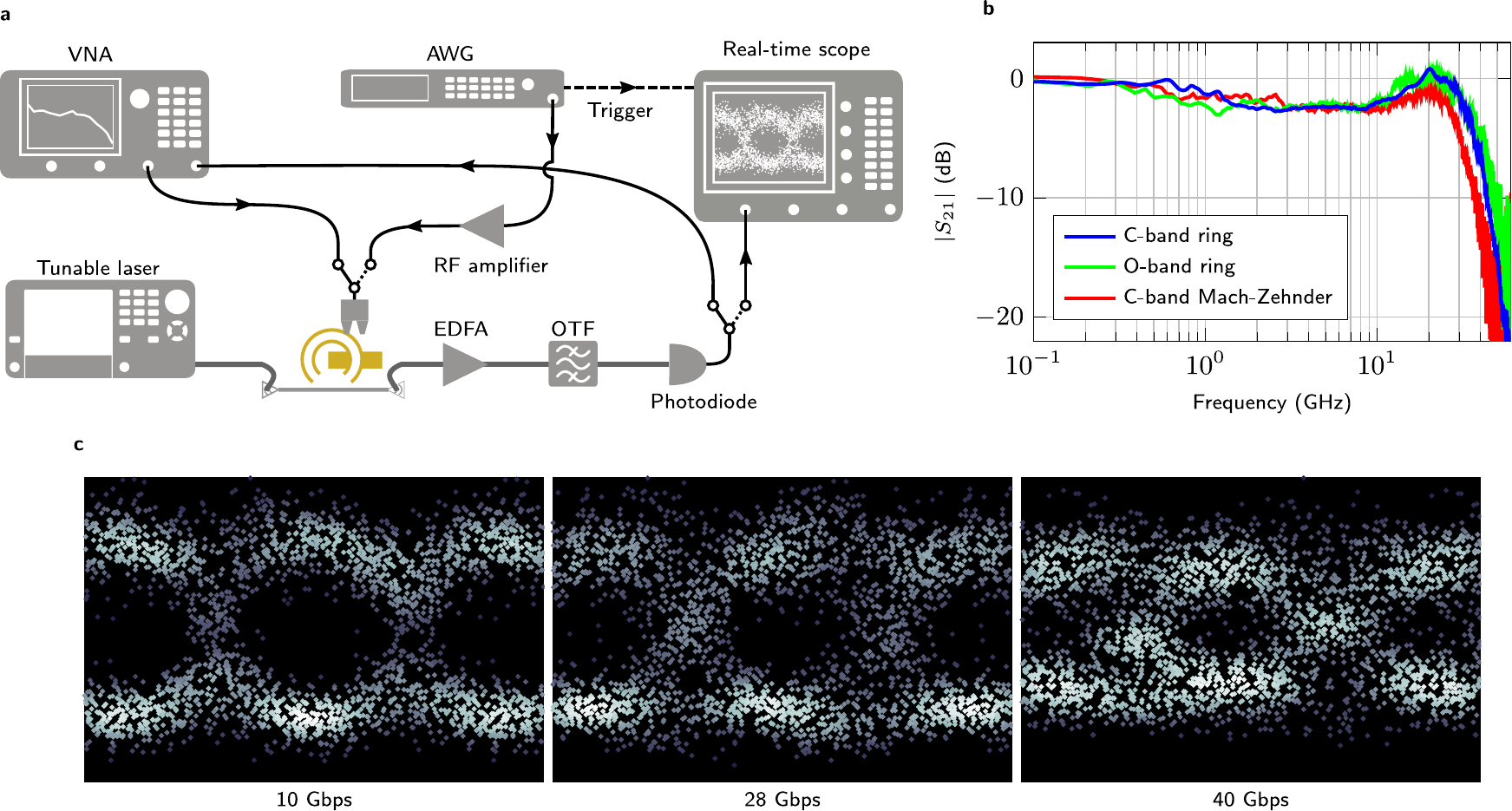}
  \caption{\sffamily\footnotesize \textbf{High-speed measurements.} \textbf{a}, Sketch of the setup used for small signal measurements (solid path in the switches) and for the eye diagram measurements (dashed path). VNA: vector network analyzer, AWG: arbitrary waveform generator, OTF: optical tunable filter. \textbf{b}, Electro-optic small signal ($\vert S_{21} \vert$ parameter) measurement of several modulators. \textbf{c}, Eye diagrams of a C-band ring modulator, measured with a non-return-to-zero scheme ($2^9-1$ pseudorandom binary sequence) and a peak-to-peak drive voltage of 4.2 V.}\label{fig:fig3}
\end{figure*}
\section*{Results}
\subsection*{Device design and fabrication}
The waveguides are patterned using 193 nm deep UV lithography in a 330 nm thick layer of LPCVD \ce{SiN} on a 3.3 \textmugreek m thick buried oxide layer, in a CMOS pilot line. Subsequently, PECVD \ce{SiO2} (thickness $\approx$ 1 \textmugreek m) is deposited over the devices and planarized, either using a combination of dry and wet etching, or by chemical-mechanical polishing (CMP). The PZT films are deposited by chemical solution deposition (CSD), using a lanthanide-based intermediate layer (see Methods and Ref. \cite{george2015lanthanide}). Finally, Ti/Au electrical contacts are patterned in the vicinity of the waveguides using photolithography, thermal evaporation and lift-off. For the samples planarized through CMP, waveguide losses of around 1 dB/cm are measured (see Supplementary Information).

Figs. \ref{fig:fig1}a and \ref{fig:fig1}b show the top view and waveguide cross-section of a C-band ring modulator, for images of the other fabricated modulators (O-band ring, C-band Mach-Zehnder), see Supplementary Information. Fig. \ref{fig:fig1}c shows a schematic of the cross-section. An electric field is applied through in-plane electrodes, changing the refractive index in the PZT and hence the effective index of the waveguide mode. The PZT thin films exhibit a higher refractive index ($n\approx$ 2.3) than \ce{SiN} ($n\approx$ 2), so a significant portion of the optical mode is confined in the PZT. A grating coupler is used for in- and outcoupling, into the fundamental quasi-TE optical mode.

\subsection*{DC characterization and poling stability}
Fig. \ref{fig:fig1}d shows the transmission spectrum of a C-band (1530 nm - 1565 nm) ring modulator. The ring has a loaded $Q$ factor of 2230 and a free-spectral range $\Delta\lambda_{\mathrm{FSR}}\approx$ 1.7 nm. The ring radius, the length of the phase shifter $L$ and the electrode spacing are respectively 100 \textmugreek m, 524 \textmugreek m and 4.4 \textmugreek m. The relatively low $Q$ factor is caused by sub-optimal alignment of the electrodes.

After deposition the PZT crystallites have one crystal plane parallel to the substrate, but no preferential orientation in the chip's plane. To obtain a significant electro-optic response for the quasi-TE optical mode, a poling step is performed by applying 60-80 V ($\approx$ 150 kV/cm) for 1 hour at room temperature, followed by several hours of stabilization time.

The transmission spectrum is measured for different DC-voltages applied across the PZT layer  (Fig. \ref{fig:fig1}e). The voltage-induced index change shifts the resonance. In Fig. \ref{fig:fig1}f, the resonance wavelength shift is plotted as a function of voltage, the slope gives the tuning efficiency $\Delta\lambda/\Delta V \approx - 13.4$ pm/V. From this we estimate the half-wave voltage-length product to be $V_\pi L =\vert L\lambda_{\mathrm{FSR}}\Delta V/(2\Delta\lambda) \vert \approx 3.3$ Vcm. Through simulation of the optical mode and DC electric field, the effective electro-optic coefficient $r_\mathrm{eff}$ of the PZT-layer is estimated to be 61 pm/V (see Methods), in good comparison with ellipsometry measurements on our thin films \cite{george2015lanthanide}. Measurements on other modulator structures yield consistent values for $r_\mathrm{eff}$, the smallest $V_\pi L$ value ($\approx 3.2$ Vcm) was measured on an O-band ring (Supplementary Information).

The PZT was poled prior to the measurements, after which no bias voltage was used. To demonstrate longer term stability of the poling, the DC tuning efficiency was periodically measured (sweeping the voltage over [-2,+2] V) on a C-band ring over a total time of almost three days. In Fig. \ref{fig:fig2}, the resulting tuning efficiency $\Delta\lambda/\Delta V$ is plotted as a function of time, decaying towards a stable value of about -13.5 pm/V over the course of several hours. The poling stabilized and there are no indications of decay over much longer periods of time, hence modulation is possible without a constant bias, as opposed to similar materials like BTO \cite{abel2013strong, xiong2014active, Eltes2017a}.

\subsection*{High-speed characterization}
For many applications, high-speed operation is essential. In Fig. \ref{fig:fig3}a the setup used for high-speed characterization is shown. On Fig. \ref{fig:fig3}b, the $|S_{21}|$ measurement for different modulators is plotted. The measured 3 dB bandwidths of both rings are around 33 GHz, the Mach-Zehnder has a bandwidth of 27 GHz. The bandwidths are not limited by the intrinsic material response of PZT, but by device design and/or characterization equipment. We furthermore demonstrate that our platform can be used for high-speed data transmission. In Fig. \ref{fig:fig3}c, eye diagrams are plotted for different bitrates, a non-return-to-zero (NRZ) binary sequence (4.2 V peak-to-peak) is used. The eye remains open up until about 40 Gbps, limited by the arbitrary waveform generator (AWG) (25 GHz bandwidth), rather than by the modulator itself. At 10 Gbps, an extinction ratio of 3.1 dB is measured (see Supplementary Information).

\begin{figure*}[!ht]
\centering
	\includegraphics{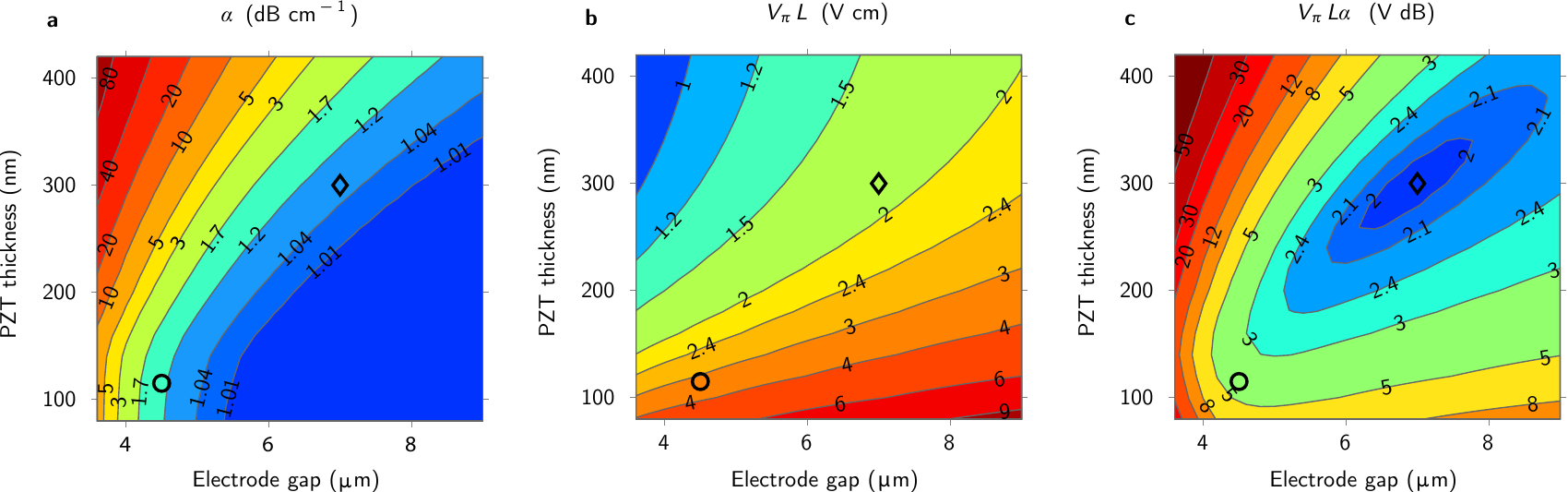}
  \caption{\sffamily\footnotesize \textbf{Numerical optimization of a PZT-on-SiN phase modulator.} Simulation of the waveguide loss $\alpha$ (\textbf{a}), the half-wave voltage-length product $V_\pi L$ (\textbf{b}) and their product  $V_\pi L \alpha$ (\textbf{c}) of a PZT-covered SiN waveguide modulator of the type shown in Fig. \ref{fig:fig1}c, for a wavelength of 1550 nm. Waveguide height, width and intermediate layer thickness are respectively 300 nm, 1.2 \textmugreek m and 20 nm. The intrinsic waveguide loss (in the absence of electrodes) was taken to be 1 dB/cm, the effective electro-optic Pockels coefficient 67 pm/V. The circles show the approximate parameters used in this work, the diamonds show the optimal point with respect to $V_\pi L \alpha$.} \label{fig:fig4}
\end{figure*}

\subsection*{Device optimization}
The presented devices were not fully optimized in terms of electro-optic modulation parameters. Primarily the PZT thickness could be increased. Sub-optimal thicknesses were used to reduce bend losses and coupling losses into PZT covered waveguide sections. These limitations can be alleviated by device design. In Fig. \ref{fig:fig4}, simulation results of the most important figures of merit are plotted as function of the PZT layer thickness, and the electrode spacing. Waveguide height, width and the wavelength are respectively 300 nm, 1.2 \textmugreek m and 1550 nm. The waveguide propagation loss $\alpha$ (Fig. \ref{fig:fig4}a) is calculated as the sum of a contribution caused by the electrodes, and a constant intrinsic propagation loss of 1 dB/cm, a realistic value if the samples are planarized using CMP (see Supplementary Information). The half-wave voltage-length product $V_\pi L$ (see Methods) and the product $V_\pi L \alpha$ are shown in Fig. \ref{fig:fig4}b and \ref{fig:fig4}c, respectively. $V_\pi L$ represents a trade-off between drive voltage and device length, $V_\pi L \alpha$  also takes into account loss, and is arguably more important for many applications \cite{Jin2018piezoelectrically}. The loss increases with decreasing electrode spacing, but also with increasing PZT thickness, since the mode expands laterally. Due to the increasing overlap between the optical mode and the PZT, $V_\pi L$ decreases with increasing thickness. $V_\pi L$ also increases with increasing electrode spacing. An optimization of the waveguide width is given in the Supplementary Information. From Fig. \ref{fig:fig4}b it is clear that $V_\pi L$ can go well below 2 Vcm. The interplay between these different dependencies can be seen in the plot of $V_\pi L \alpha$ (Fig. \ref{fig:fig4}b), which shows an optimum with $V_\pi L \alpha \approx 2$ VdB.

\section*{Discussion}
To conclude, we have demonstrated a novel platform for efficient, optically broadband, high-speed, nano-photonic electro-optic modulators. Using a relatively simple chemical solution deposition procedure we incorporated a thin film of strongly electro-optic PZT onto a SiN-based photonic chip. We demonstrated stable poling of the electro-optic material, and efficient and high-speed modulation, in the absence of a bias voltage. O- and C-band operation was shown, however we expect the platform to be operational into the visible wavelength range ($>$450 nm) \cite{rahim2017expanding, pandey2005structural, gu2017lanthanum}. From simulations it is clear that the devices characterized in this paper do not yet represent the limitations of the platform and $V_\pi L \alpha \approx 2$ VdB is achievable. Moreover, our approach is unique in its versatility, as the PZT film can be deposited on any sufficiently flat surface, enabling the incorporation of the electro-optic films onto other guided-wave platforms.

\section*{Methods}
\subsection*{PZT deposition and patterning.}
The details of the lanthanide-assisted deposition procedure have been published elsewhere \cite{george2015lanthanide}, a short summary is given here. Intermediate seed layers based on lanthanides are deposited prior to the PZT deposition. The intermediate layer acts as a barrier layer to prevent the inter-diffusion of elements and as a seed layer providing the lattice match to grow highly oriented thin films. A critical thickness of the intermediate layer needs to be maintained ($>$ 5 nm) to avoid diffusion and secondary phase formations. However, on samples with considerable surface topology, thicker intermediate layers are necessary to provide good step coverage and to avoid any issues associated with the conformity in spin-coating. On our samples planarized through etching, step heights between oxide and \ce{SiN} waveguides varied. We typically used an intermediate layer of thickness $\approx$ 24 nm to avoid issues. Both the intermediate layer and the PZT thin films are deposited by repeating the spin-coating and annealing procedure, which allows easy control of the film thickness. The PZT layer is deposited and annealed at 620 $^{\circ}$C for 15 min in tube furnace under an oxygen ambient. This Chemical Solution Deposition (CSD) method, also called sol-gel, provides a cheap and flexible alternative to achieve high quality stoichiometric PZT thin films regardless the substrate material. A reactive ion etching (RIE) procedure based on \ce{SF6} chemistry is used to pattern the PZT layer. The PZT film was removed selectively over the grating couplers used for the optical measurements.

\subsection*{High-speed measurements.}
The small-signal response measurements were performed using an Agilent PNA-X N5247A network analyzer and a high-speed photodiode (Discovery Semiconductors DSC10H Optical Receiver). For the eye diagram measurements, an arbitrary waveform generator (Keysight AWG M8195A) and RF amplifier (SHF S807) are used to apply a pseudorandom non-return-to-zero (NRZ) binary sequence, the modulator output is measured with a Keysight 86100D oscilloscope with 50 GHz bandwidth and Discovery Semiconductors DSC-R409 PIN-TIA Optical Receiver.

\subsection*{Calculation of the electro-optic parameters.}
Using COMSOL Multiphysics\textsuperscript{\textregistered}, several parameters can be calculated that strongly influence the performance of the modulators. To obtain efficient phase modulation, it is essential to maximize the overlap between the optical mode and the RF electrical signal, quantified by the electro-optic overlap integral \cite{koeber2015femtojoule},
\begin{align}
\Gamma = \frac{g}{V} \frac{\epsilon_0 c n_{\mathrm{PZT}}\iint_{\mathrm{PZT}} E^\mathrm{e}_x \vert E^{\mathrm{op}}_x\vert^2 \mathrm{d}x\mathrm{d}y}{\iint \mathrm{Re}(\mathbf{E^{\mathrm{op}}}\times\mathbf{H^{\mathrm{op^*}}})\cdot \mathrm{\hat{e}}_z \mathrm{d}x\mathrm{d}y},
\end{align}
where $g$ is the spacing between the electrodes, $V$ the applied voltage, $\epsilon_0$ the vacuum permittivity, $c$ the speed of light in vacuum and $n_{\mathrm{PZT}}$ the refractive index of PZT. $E^\mathrm{e}_x$ is the in-plane (x-)component of the RF electric field, and $E^{\mathrm{op}}_x$ represents the in-plane transversal component of the optical field. When used as a phase shifter, an important figure of merit is the half-wave voltage-length product $V_\pi L$. This product relates to the electro-optic coefficient $r_{\mathrm{eff}}$ of the PZT films and to $\Gamma$ \cite{koeber2015femtojoule},
\begin{align}
 V_\pi L = \frac{\lambda g}{n_{\mathrm{PZT}}^3 \Gamma r_{\mathrm{eff}}},
 \label{eq:VpiL}
\end{align}
where $\lambda$ is the wavelength. Another important parameter is the propagation loss of the optical mode, consisting of an intrinsic contribution (scattering, material loss in the PZT, intermediate layer, nitride and oxide) and a contribution caused by the vicinity of the electrical contacts. The former can be estimated based on cut-back measurements on unmetalized waveguides (see Supplementary Information), the latter can be numerically calculated.\\

\section*{Acknowledgements}
The authors thank St\'ephane Clemmen for his overseeing role in the \ce{SiN} chip fabrication and Philippe F. Smet for help with processing. K. A. is funded by FWO Flanders.

\bibliographystyle{naturemag}

\begin{thebibliography}{10}
\expandafter\ifx\csname url\endcsname\relax
  \def\url#1{\texttt{#1}}\fi
\expandafter\ifx\csname urlprefix\endcsname\relax\def\urlprefix{URL }\fi
\providecommand{\bibinfo}[2]{#2}
\providecommand{\eprint}[2][]{\url{#2}}

\bibitem{Sun2015single}
\bibinfo{author}{Sun, C.} \emph{et~al.}
\newblock \bibinfo{title}{Single-chip microprocessor that communicates directly
  using light}.
\newblock \emph{\bibinfo{journal}{Nature}} \textbf{\bibinfo{volume}{528}},
  \bibinfo{pages}{534-538} (\bibinfo{year}{2015}).

\bibitem{rahim2017expanding}
\bibinfo{author}{Rahim, A.} \emph{et~al.}
\newblock \bibinfo{title}{Expanding the silicon photonics portfolio with
  silicon nitride photonic integrated circuits}.
\newblock \emph{\bibinfo{journal}{Journal of Lightwave Technology}}
  \textbf{\bibinfo{volume}{35}}, \bibinfo{pages}{639--649}
  (\bibinfo{year}{2017}).

\bibitem{levy2010cmos}
\bibinfo{author}{Levy, J.~S.} \emph{et~al.}
\newblock \bibinfo{title}{{CMOS}-compatible multiple-wavelength oscillator for
  on-chip optical interconnects}.
\newblock \emph{\bibinfo{journal}{Nature Photonics}}
  \textbf{\bibinfo{volume}{4}}, \bibinfo{pages}{37} (\bibinfo{year}{2010}).

\bibitem{bauters2011ultra}
\bibinfo{author}{Bauters, J.~F.} \emph{et~al.}
\newblock \bibinfo{title}{Ultra-low-loss high-aspect-ratio \ce{Si3N4}
  waveguides}.
\newblock \emph{\bibinfo{journal}{Optics Express}}
  \textbf{\bibinfo{volume}{19}}, \bibinfo{pages}{3163--3174}
  (\bibinfo{year}{2011}).

\bibitem{moss2013new}
\bibinfo{author}{Moss, D.~J.}, \bibinfo{author}{Morandotti, R.},
  \bibinfo{author}{Gaeta, A.~L.} \& \bibinfo{author}{Lipson, M.}
\newblock \bibinfo{title}{New {CMOS}-compatible platforms based on silicon
  nitride and {Hydex} for nonlinear optics}.
\newblock \emph{\bibinfo{journal}{Nature Photonics}}
  \textbf{\bibinfo{volume}{7}}, \bibinfo{pages}{597} (\bibinfo{year}{2013}).

\bibitem{ramelow2015silicon}
\bibinfo{author}{Ramelow, S.} \emph{et~al.}
\newblock \bibinfo{title}{Silicon-nitride platform for narrowband entangled
  photon generation}.
\newblock \emph{\bibinfo{journal}{arXiv preprint arXiv:1508.04358}}
  (\bibinfo{year}{2015}).

\bibitem{kahl2015waveguide}
\bibinfo{author}{Kahl, O.} \emph{et~al.}
\newblock \bibinfo{title}{Waveguide integrated superconducting single-photon
  detectors with high internal quantum efficiency at telecom wavelengths}.
\newblock \emph{\bibinfo{journal}{Scientific Reports}}
  \textbf{\bibinfo{volume}{5}}, \bibinfo{pages}{10941} (\bibinfo{year}{2015}).

\bibitem{zhuang2015programmable}
\bibinfo{author}{Zhuang, L.}, \bibinfo{author}{Roeloffzen, C.~G.},
  \bibinfo{author}{Hoekman, M.}, \bibinfo{author}{Boller, K.-J.} \&
  \bibinfo{author}{Lowery, A.~J.}
\newblock \bibinfo{title}{Programmable photonic signal processor chip for
  radiofrequency applications}.
\newblock \emph{\bibinfo{journal}{Optica}} \textbf{\bibinfo{volume}{2}},
  \bibinfo{pages}{854--859} (\bibinfo{year}{2015}).

\bibitem{poulton2017large}
\bibinfo{author}{Poulton, C.~V.} \emph{et~al.}
\newblock \bibinfo{title}{Large-scale silicon nitride nanophotonic phased
  arrays at infrared and visible wavelengths}.
\newblock \emph{\bibinfo{journal}{Optics Letters}}
  \textbf{\bibinfo{volume}{42}}, \bibinfo{pages}{21--24}
  (\bibinfo{year}{2017}).

\bibitem{reed2014high}
\bibinfo{author}{Reed, G.~T.} \emph{et~al.}
\newblock \bibinfo{title}{High-speed carrier-depletion silicon {Mach-Zehnder}
  optical modulators with lateral {PN} junctions}.
\newblock \emph{\bibinfo{journal}{Frontiers in Physics}}
  \textbf{\bibinfo{volume}{2}}, \bibinfo{pages}{77} (\bibinfo{year}{2014}).

\bibitem{xu2005micrometre}
\bibinfo{author}{Xu, Q.}, \bibinfo{author}{Schmidt, B.},
  \bibinfo{author}{Pradhan, S.} \& \bibinfo{author}{Lipson, M.}
\newblock \bibinfo{title}{Micrometre-scale silicon electro-optic modulator}.
\newblock \emph{\bibinfo{journal}{Nature}} \textbf{\bibinfo{volume}{435}},
  \bibinfo{pages}{325} (\bibinfo{year}{2005}).

\bibitem{liu2004high}
\bibinfo{author}{Liu, A.} \emph{et~al.}
\newblock \bibinfo{title}{A high-speed silicon optical modulator based on a
  metal--oxide--semiconductor capacitor}.
\newblock \emph{\bibinfo{journal}{Nature}} \textbf{\bibinfo{volume}{427}},
  \bibinfo{pages}{615} (\bibinfo{year}{2004}).

\bibitem{hiraki2017heterogeneously}
\bibinfo{author}{Hiraki, T.} \emph{et~al.}
\newblock \bibinfo{title}{Heterogeneously integrated {III--V/Si MOS} capacitor
  {Mach--Zehnder} modulator}.
\newblock \emph{\bibinfo{journal}{Nature Photonics}}
  \textbf{\bibinfo{volume}{11}}, \bibinfo{pages}{482} (\bibinfo{year}{2017}).

\bibitem{han2017efficient}
\bibinfo{author}{Han, J.-H.} \emph{et~al.}
\newblock \bibinfo{title}{Efficient low-loss {InGaAsP/Si} hybrid {MOS} optical
  modulator}.
\newblock \emph{\bibinfo{journal}{Nature Photonics}}
  \textbf{\bibinfo{volume}{11}}, \bibinfo{pages}{486} (\bibinfo{year}{2017}).

\bibitem{liu2011graphene}
\bibinfo{author}{Liu, M.} \emph{et~al.}
\newblock \bibinfo{title}{A graphene-based broadband optical modulator}.
\newblock \emph{\bibinfo{journal}{Nature}} \textbf{\bibinfo{volume}{474}},
  \bibinfo{pages}{64} (\bibinfo{year}{2011}).

\bibitem{sorianello2018graphene}
\bibinfo{author}{Sorianello, V.} \emph{et~al.}
\newblock \bibinfo{title}{Graphene--silicon phase modulators with gigahertz
  bandwidth}.
\newblock \emph{\bibinfo{journal}{Nature Photonics}}
  \textbf{\bibinfo{volume}{12}}, \bibinfo{pages}{40} (\bibinfo{year}{2018}).

\bibitem{alloatti2014100}
\bibinfo{author}{Alloatti, L.} \emph{et~al.}
\newblock \bibinfo{title}{100 {GHz} silicon--organic hybrid modulator}.
\newblock \emph{\bibinfo{journal}{Light: Science \& Applications}}
  \textbf{\bibinfo{volume}{3}}, \bibinfo{pages}{e173} (\bibinfo{year}{2014}).

\bibitem{srinivasan201656}
\bibinfo{author}{Srinivasan, A.} \emph{et~al.}
\newblock \bibinfo{title}{56 {Gb/s} germanium waveguide electro-absorption
  modulator}.
\newblock \emph{\bibinfo{journal}{Journal of Lightwave Technology}}
  \textbf{\bibinfo{volume}{34}}, \bibinfo{pages}{419--424}
  (\bibinfo{year}{2016}).

\bibitem{abel2013strong}
\bibinfo{author}{Abel, S.} \emph{et~al.}
\newblock \bibinfo{title}{A strong electro-optically active lead-free
  ferroelectric integrated on silicon}.
\newblock \emph{\bibinfo{journal}{Nature Communications}}
  \textbf{\bibinfo{volume}{4}}, \bibinfo{pages}{1671} (\bibinfo{year}{2013}).

\bibitem{xiong2014active}
\bibinfo{author}{Xiong, C.} \emph{et~al.}
\newblock \bibinfo{title}{Active silicon integrated nanophotonics:
  ferroelectric \ce{BaTiO3}-devices}.
\newblock \emph{\bibinfo{journal}{Nano Letters}} \textbf{\bibinfo{volume}{14}},
  \bibinfo{pages}{1419--1425} (\bibinfo{year}{2014}).

\bibitem{Eltes2017a}
\bibinfo{author}{Eltes, F.} \emph{et~al.}
\newblock \bibinfo{title}{A novel 25 {Gbps} electro-optic {Pockels} modulator
  integrated on an advanced \ce{Si} photonic platform}.
\newblock \emph{\bibinfo{journal}{IEEE International Electron Devices Meeting
  (IEDM)}}  (\bibinfo{year}{2017}).

\bibitem{xiong2012low}
\bibinfo{author}{Xiong, C.}, \bibinfo{author}{Pernice, W.~H.} \&
  \bibinfo{author}{Tang, H.~X.}
\newblock \bibinfo{title}{Low-loss, silicon integrated, aluminum nitride
  photonic circuits and their use for electro-optic signal processing}.
\newblock \emph{\bibinfo{journal}{Nano Letters}} \textbf{\bibinfo{volume}{12}},
  \bibinfo{pages}{3562--3568} (\bibinfo{year}{2012}).

\bibitem{wang2018nanophotonic}
\bibinfo{author}{Wang, C.}, \bibinfo{author}{Zhang, M.},
  \bibinfo{author}{Stern, B.}, \bibinfo{author}{Lipson, M.} \&
  \bibinfo{author}{Lon{\v{c}}ar, M.}
\newblock \bibinfo{title}{Nanophotonic lithium niobate electro-optic
  modulators}.
\newblock \emph{\bibinfo{journal}{Optics Express}}
  \textbf{\bibinfo{volume}{26}}, \bibinfo{pages}{1547--1555}
  (\bibinfo{year}{2018}).

\bibitem{phare2015graphene}
\bibinfo{author}{Phare, C.~T.}, \bibinfo{author}{Lee, Y.-H.~D.},
  \bibinfo{author}{Cardenas, J.} \& \bibinfo{author}{Lipson, M.}
\newblock \bibinfo{title}{Graphene electro-optic modulator with 30 {GHz}
  bandwidth}.
\newblock \emph{\bibinfo{journal}{Nature Photonics}}
  \textbf{\bibinfo{volume}{9}}, \bibinfo{pages}{511--514}
  (\bibinfo{year}{2015}).

\bibitem{hosseini2015stress}
\bibinfo{author}{Hosseini, N.} \emph{et~al.}
\newblock \bibinfo{title}{Stress-optic modulator in {TriPleX} platform using a
  piezoelectric lead zirconate titanate ({PZT}) thin film}.
\newblock \emph{\bibinfo{journal}{Optics Express}}
  \textbf{\bibinfo{volume}{23}}, \bibinfo{pages}{14018--14026}
  (\bibinfo{year}{2015}).

\bibitem{Jin2018piezoelectrically}
\bibinfo{author}{Jin, W.}, \bibinfo{author}{Polcawich, R.~G.},
  \bibinfo{author}{Morton, P.~A.} \& \bibinfo{author}{Bowers, J.~E.}
\newblock \bibinfo{title}{Piezoelectrically tuned silicon nitride ring
  resonator}.
\newblock \emph{\bibinfo{journal}{Optics Express}}
  \textbf{\bibinfo{volume}{26}}, \bibinfo{pages}{3174--3187}
  (\bibinfo{year}{2018}).

\bibitem{george2015lanthanide}
\bibinfo{author}{George, J.} \emph{et~al.}
\newblock \bibinfo{title}{Lanthanide-assisted deposition of strongly
  electro-optic {PZT} thin films on silicon: toward integrated active
  nanophotonic devices}.
\newblock \emph{\bibinfo{journal}{ACS Applied Materials \& Interfaces}}
  \textbf{\bibinfo{volume}{7}}, \bibinfo{pages}{13350--13359}
  (\bibinfo{year}{2015}).

\bibitem{reed2014recent}
\bibinfo{author}{Reed, G.~T.} \emph{et~al.}
\newblock \bibinfo{title}{Recent breakthroughs in carrier depletion based
  silicon optical modulators}.
\newblock \emph{\bibinfo{journal}{Nanophotonics}} \textbf{\bibinfo{volume}{3}},
  \bibinfo{pages}{229--245} (\bibinfo{year}{2014}).

\bibitem{pandey2005structural}
\bibinfo{author}{Pandey, S.} \emph{et~al.}
\newblock \bibinfo{title}{Structural, ferroelectric and optical properties of
  {PZT} thin films}.
\newblock \emph{\bibinfo{journal}{Physica B: Condensed Matter}}
  \textbf{\bibinfo{volume}{369}}, \bibinfo{pages}{135--142}
  (\bibinfo{year}{2005}).

\bibitem{gu2017lanthanum}
\bibinfo{author}{Gu, W.}, \bibinfo{author}{Song, Y.}, \bibinfo{author}{Liu, J.}
  \& \bibinfo{author}{Wang, F.}
\newblock \bibinfo{title}{Lanthanum-based compounds: Electronic
  bandgap-dependent electro-catalytic materials toward oxygen reduction
  reaction}.
\newblock \emph{\bibinfo{journal}{Chemistry-A European Journal}}
  (\bibinfo{year}{2017}).
  
  \bibitem{koeber2015femtojoule}
\bibinfo{author}{Koeber, S.} \emph{et~al.}
\newblock \bibinfo{title}{Femtojoule electro-optic modulation using a
  silicon-organic hybrid device}.
\newblock \emph{\bibinfo{journal}{Light: Science \& Applications}}
  \textbf{\bibinfo{volume}{4}}, \bibinfo{pages}{e255} (\bibinfo{year}{2015}).

\end{thebibliography}

\clearpage

\onecolumngrid
\setcounter{figure}{0}
\renewcommand{\thefigure}{S\arabic{figure}}

\linespread{1.5}

\section*{Supplementary information}
\setcounter{subsection}{0}
\renewcommand{\thesubsection}{S\arabic{subsection}}

\subsection{Additional devices}
\begin{figure*}[h]
  \includegraphics[width=\textwidth]{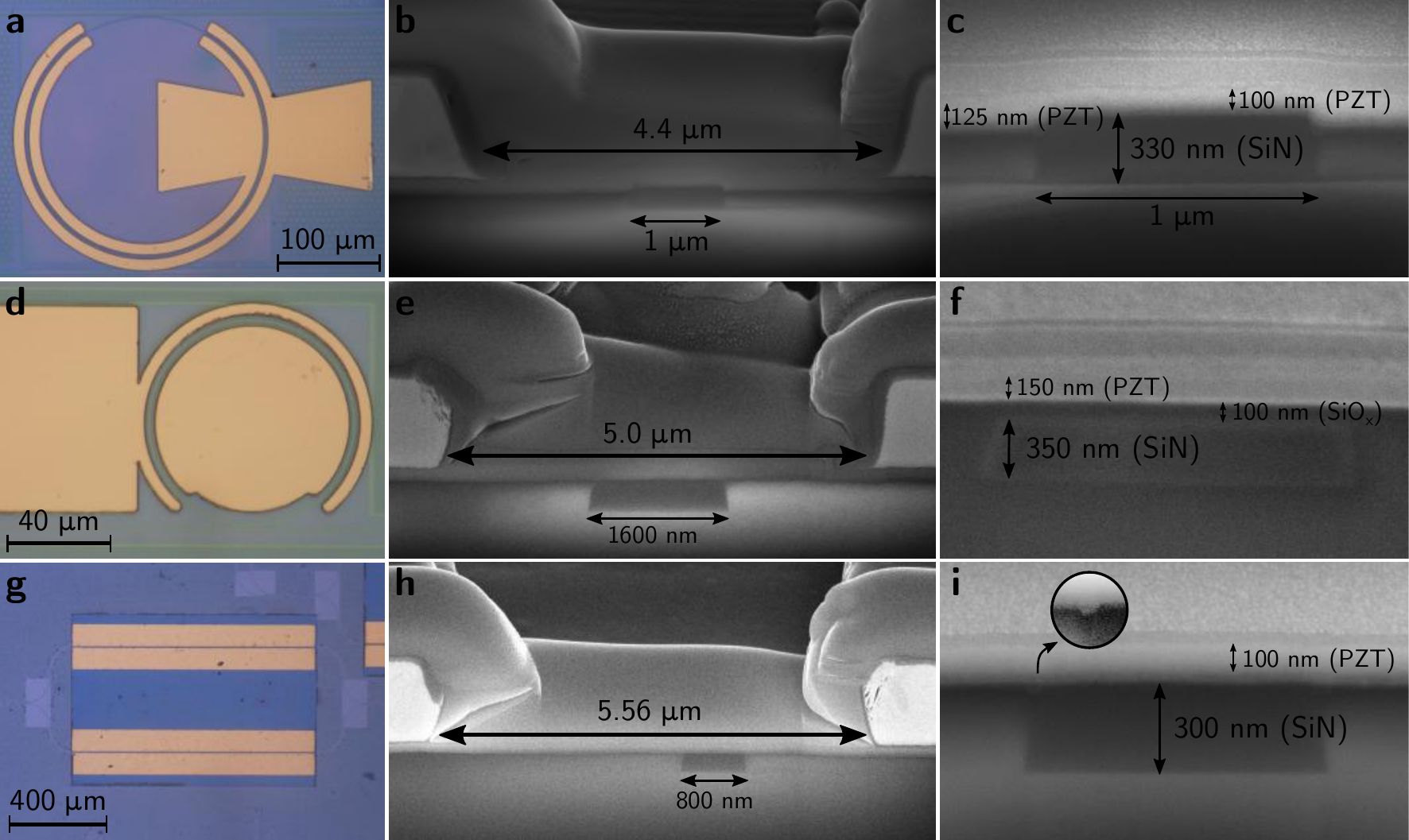}
    \caption{\sffamily\footnotesize \textbf{Optical microscope and SEM images of the different modulator types.} \textbf{a}, Top view of a C-band ring modulator. \textbf{b}, \textbf{c}, Cross-sections of a C-band ring modulator. \textbf{d}, Top view of a O-band ring modulator. \textbf{e}, \textbf{f}, Cross-section of a O-band ring modulator. \textbf{g}, Top view of a C-band MZI modulator. \textbf{h}, \textbf{i}, Cross-section of a C-band MZI modulator, the inset shows the trench at the waveguide edge with enhanced contrast, caused by nonuniform etch rates. The nominal thickness of the intermediate lanthanide layer (below the PZT) is 24 nm in all devices.}\label{fig:microscope_FIB}
\end{figure*}

\begin{figure*}[h!]
\centering
	\includegraphics{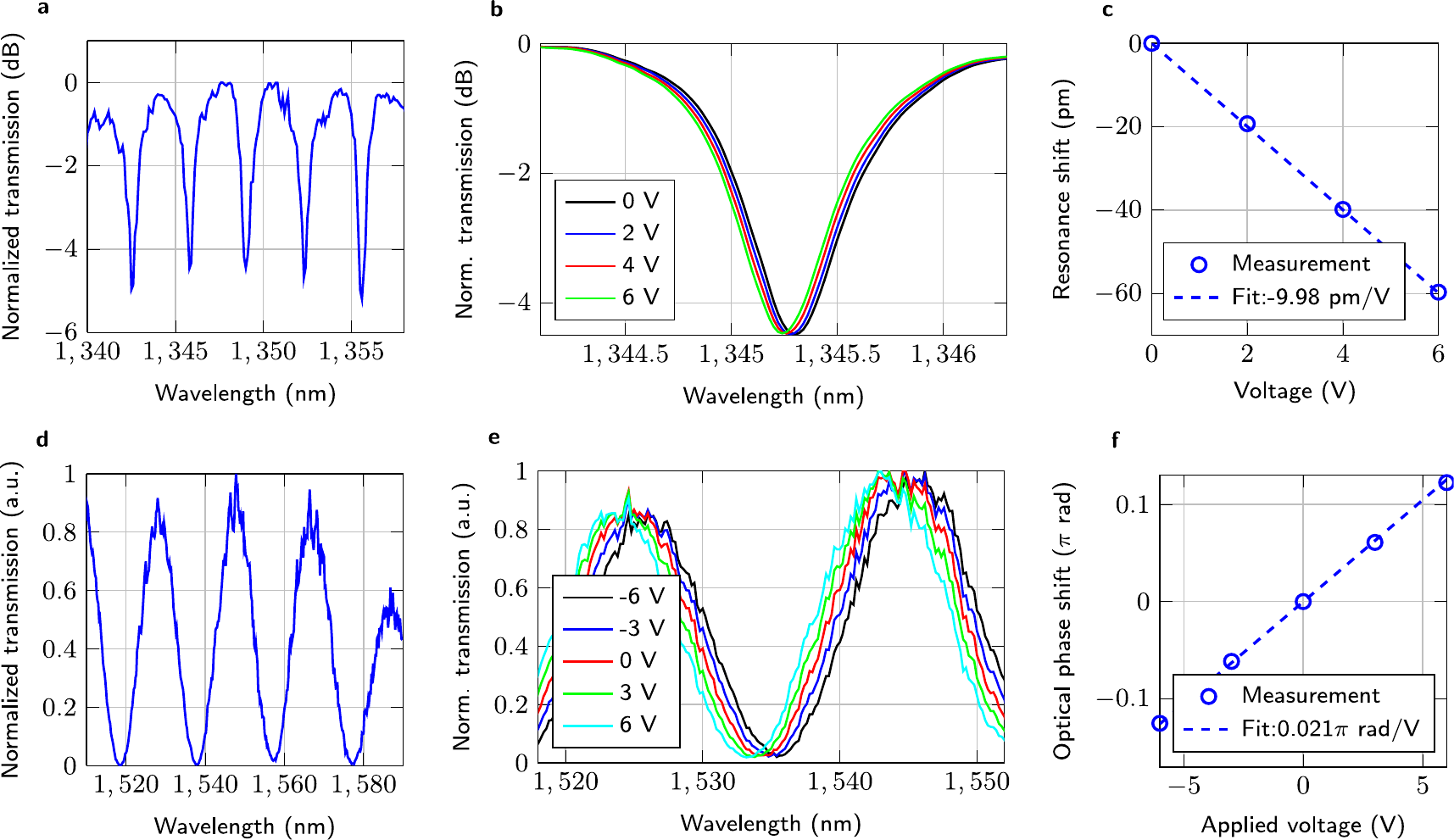}
  \caption{\sffamily\footnotesize \textbf{Static response of O-band ring and C-band Mach-Zehnder modulators.} \textbf{a}, Normalized transmission spectrum of the O-band ring. \textbf{b}, Transmission spectra for different DC voltages. \textbf{c}, Resonance wavelength shift versus voltage, including a linear fit. \textbf{d}, Normalized transmission spectrum of the C-band MZI. \textbf{e}, Transmission spectra for different DC voltages. \textbf{f}, Optical phase shift as a function of voltage, including linear fit.}\label{fig:transmission_shift_all}
\end{figure*}

The paper mainly highlights the results on C-band ring resonator modulators. However, more device types have been fabricated and characterized. Apart from the C-band rings, this section provides more details on O-band (1260 nm -  1360 nm)  ring modulators and a C-band Mach-Zehnder type modulator. Fig. \ref{fig:microscope_FIB} gives an overview: Figs. \ref{fig:microscope_FIB}a-c show respectively the top-view, cross section and a detailed cross section of the C-band device described in the paper, whereas Figs. \ref{fig:microscope_FIB}d-f and \ref{fig:microscope_FIB}g-i do the same for an O-band ring and a C-band Mach-Zehnder Interferometer (MZI). The O-band ring has a $Q$-factor of 1820 and a free-spectral range $\Delta\lambda_{\mathrm{FSR}}$ of 3.27 nm. The ring radius is 40 \textmugreek m, with a phase shifter length $L$ of 195 \textmugreek m. The MZI modulator electrodes have a length of 1 mm. Note the differences between the cross-sections of the devices, the waveguides in Figs. \ref{fig:microscope_FIB}c and \ref{fig:microscope_FIB}i were planarized trough back-etching of the top ($\approx$ 1 \textmugreek m thick) oxide using a combination of reactive ion etching (RIE) and wet etching (HF), variations in top oxide thickness, etch rates and exact etch times can lead to relatively large steps (Fig. \ref{fig:microscope_FIB}c). Moreover, the etch rates of the deposited oxide depend on the exact nitride structures underneath, even in the case of a seemingly planar surface (Fig. \ref{fig:microscope_FIB}i), trenches of several tens of nanometers arise next to the waveguide (see inset in Fig. \ref{fig:microscope_FIB}i). On Fig. \ref{fig:microscope_FIB}d-f however, a device planarized using chemical mechanical polishing (CMP) is shown. A buffer layer of 50-100 nm of oxide is left on top of the nitride waveguide, so the obtained surface is much smoother. This leads to much smaller propagation losses (see section S2).

Figs. \ref{fig:transmission_shift_all}a-c respectively show the transmission spectrum of the O-band ring (pictures in Figs. \ref{fig:microscope_FIB}d-f), its transmission spectrum for different DC-voltages, and the resonance shift as a function of voltage. The linear fit on Fig. \ref{fig:transmission_shift_all}c shows a resonance tuning efficiency of $\Delta\lambda/\Delta V \approx -10$ pm/V. From this value the half-wave voltage-length product can be estimated:  $V_\pi L =\vert L\lambda_{\mathrm{FSR}}\Delta V/(2\Delta\lambda) \vert\approx 3.19$ Vcm. Through simulation of the optical mode and DC electric field, the effective electro-optic coefficient  $r_{\mathrm{eff}}$ of the PZT-layer around 1310 nm is estimated to be about 67 pm/V (see Methods).\\
Figs. \ref{fig:transmission_shift_all}d-f show the transmission spectrum of the MZI modulator (pictures in Figs. \ref{fig:microscope_FIB}g-i), the transmission spectrum for different voltages applied across the PZT and the electro-optic phase shift (with respect to 0 V) as a function of voltage. The voltage is applied to only one of the MZI arms. From this, we can estimate the $V_\pi$ (voltage needed to induce a $\pi$ phase shift, or a shift of the sinusoidal transmission pattern over half a period) to be 47.6 V, corresponding to a $V_\pi L$ of 4.76 Vcm. This corresponds to an  $r_{\mathrm{eff}}$ of the PZT-layer of about 70 pm/V.

Variations in the measured $V_\pi L$ values are mainly due to variations in the waveguide cross-sections, electrode spacings and the used wavelengths (C-band versus O-band), see Eq.  \eqref{eq:VpiL}. Extracted electro-optic coefficients $r_{\mathrm{eff}}$ also vary somewhat, differences can in part be due to variations in film quality on different samples, but mainly stem from small uncertainties on the exact cross-section dimensions.
\clearpage
\subsection{Waveguide loss measurements}\label{sec:loss}
\begin{figure*}[h]
\centering
	\includegraphics{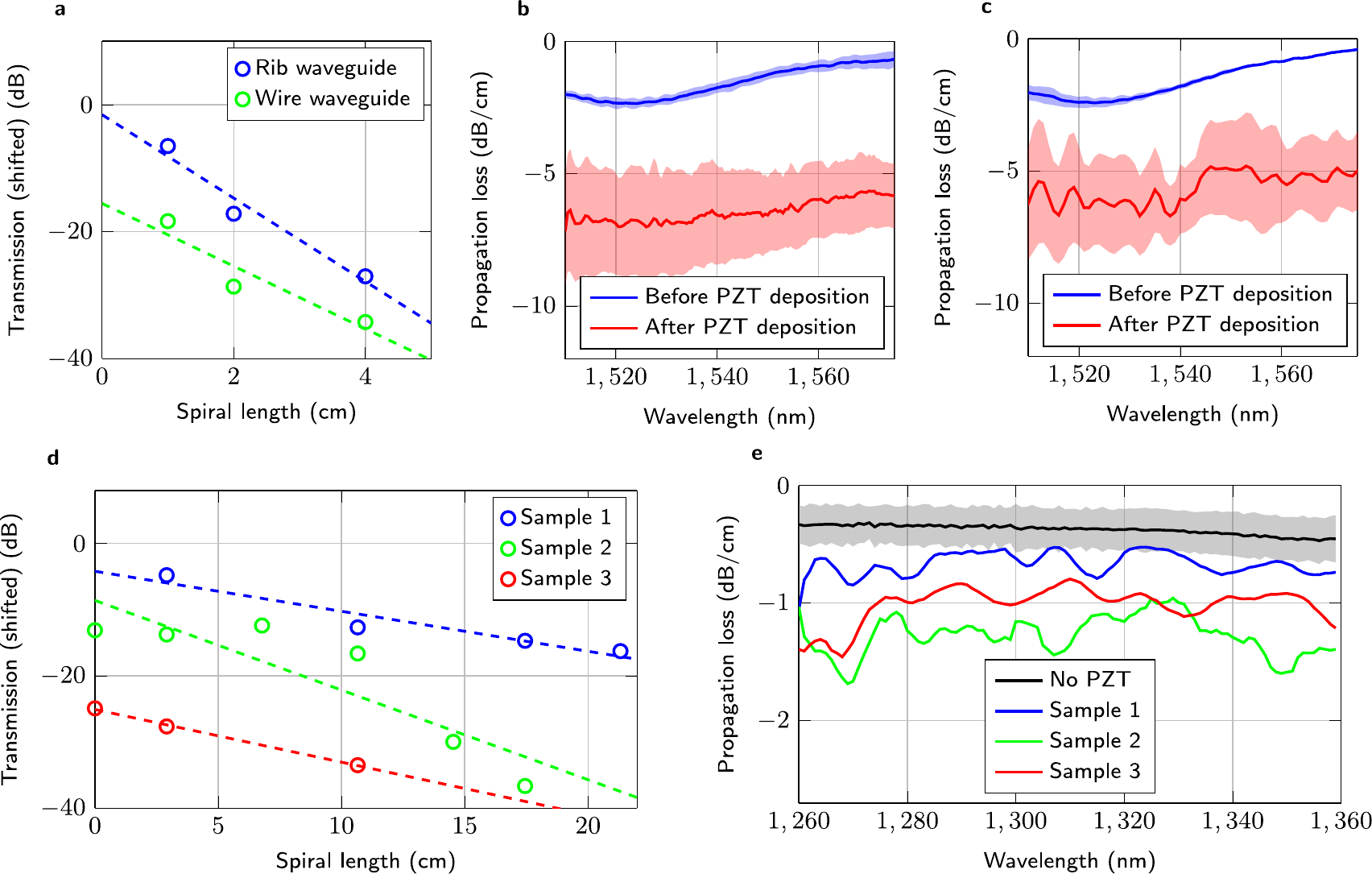}
	\caption{\sffamily\footnotesize\textbf{Loss measurements.} \textbf{a}, Transmission versus waveguide length for a PZT-covered rib and wire waveguide, with cross-section similar to Fig. \ref{fig:microscope_FIB}c (width =  1400 nm, wavelength = 1550 nm, PZT thickness $\approx$ 125 nm). \textbf{b}, \textbf{c}, Propagation loss of the respective rib and wire waveguides versus wavelength, before and after PZT deposition. The shaded area shows the standard deviation on the fitted slope. \textbf{d}, Transmission versus waveguide length for a PZT-covered waveguide, with cross-section similar to Fig. \ref{fig:microscope_FIB}f (width =  800 nm, wavelength = 1310 nm, PZT thickness = 150 nm). Measured on 3 different samples. \textbf{e}, Propagation loss versus wavelength for these waveguide sets, including a sample with no PZT.}\label{fig:figS3}
\end{figure*}

In Fig. \ref{fig:figS3} the loss measurements on different types of waveguides are summarized. For the C-band measurements, chips were planarized using a combination of reactive ion etching (RIE) and wet hydrogen fluoride (HF) etching. Typically resulting in steps and trenches of several tens of nanometers in the vicinity of the waveguide (see section S1 and Fig. \ref{fig:microscope_FIB}c, i). Figs. \ref{fig:figS3}a-c summarize loss measurements on such waveguides, for a set of rib waveguide spirals (blue line on Fig. \ref{fig:figS3}a and Fig. \ref{fig:figS3}b) and a set of wire waveguide spirals (green line on Fig. \ref{fig:figS3}a and Fig. \ref{fig:figS3}c). The PZT-covered wire waveguides, resembling the ones used in the C-band modulators, have an estimated loss of 5 to 6 dB/cm. The rib waveguides were defined using a partial etch of 220 nm next to the waveguide core, the influence of this on the propagation loss is only expected to be minor, as is demonstrated by the measurements. Note that before PZT deposition, the waveguide loss can be as low as 0.5 dB/cm. For the O-band measurements, the planarization of the waveguides was done by chemical-mechanical polishing (CMP), resulting in a waveguide cross-section as shown in Fig. \ref{fig:microscope_FIB}f, with a residual oxide layer of 50 to 100 nm on top of the waveguide (see section S1). Figs. \ref{fig:figS3}d-e show the loss measurements of such waveguides for 3 test samples. The smoother surface for the PZT deposition can result in losses below 1 dB/cm. The simulated confinement factor in the PZT layer for the C- and O-band waveguides used in the loss measurements are respectively $\approx0.23$ (for both rib and wire waveguides) and  $\approx0.3$.
\clearpage
\subsection{Extinction ratio measurement}
\begin{figure*}[!ht]
\centering
	\includegraphics{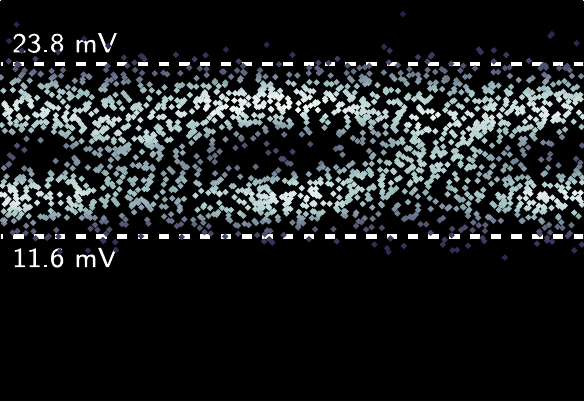}
  \caption{\sffamily\footnotesize \textbf{Extinction ratio measurement.}  Eye diagram of a C-band ring modulator, measured with a 10Gbps non-return-to-zero scheme and a peak-to-peak voltage of 4.2 V (same as in Fig. \ref{fig:fig2}c). Obtained using a DC-coupled optical receiver.}\label{fig:figS4}
\end{figure*}

The eye diagram shown in Fig. \ref{fig:figS4} was obtained using a DC-coupled Tectronix 80 C02-CR optical receiver with a sampling oscilloscope (Tektronix CSA 8000), applying a peak-to-peak voltage of 4.2 V at 10 Gbps (same as for Fig. \ref{fig:fig2}c). Since the measured voltage scales with the total optical power, we can estimate the extinction ratio to be about $10\cdot \log_{10}(P_{\mathrm{max}}/P_{\mathrm{min}})\;\mathrm{dB} \approx\log_{10}(23.8/11.6)\;\mathrm{dB} = 3.12\;\mathrm{dB}$. This corresponds well with a simple ball-park estimate based on the observed transmission spectrum and static DC-shift (Figs. \ref{fig:fig1}d, e), since the extinction ratio in DC can be estimated as $\Delta T \approx \vert \left[\frac{dT}{d\lambda}\right]_{\mathrm{max}} \cdot \frac{d\lambda}{dV} \cdot V_{\mathrm{pp}} \vert \approx 60\;\mathrm{dB/nm}\cdot 13.5 \;\mathrm{pm/V} \cdot 4.2 \;\mathrm{V} = 3.4 \;\mathrm{dB}$, where $T$ is the transmission expressed in dB.
\clearpage
\subsection{Device optimization - influence of the waveguide width}
\begin{figure*}[!ht]
\centering
	\includegraphics{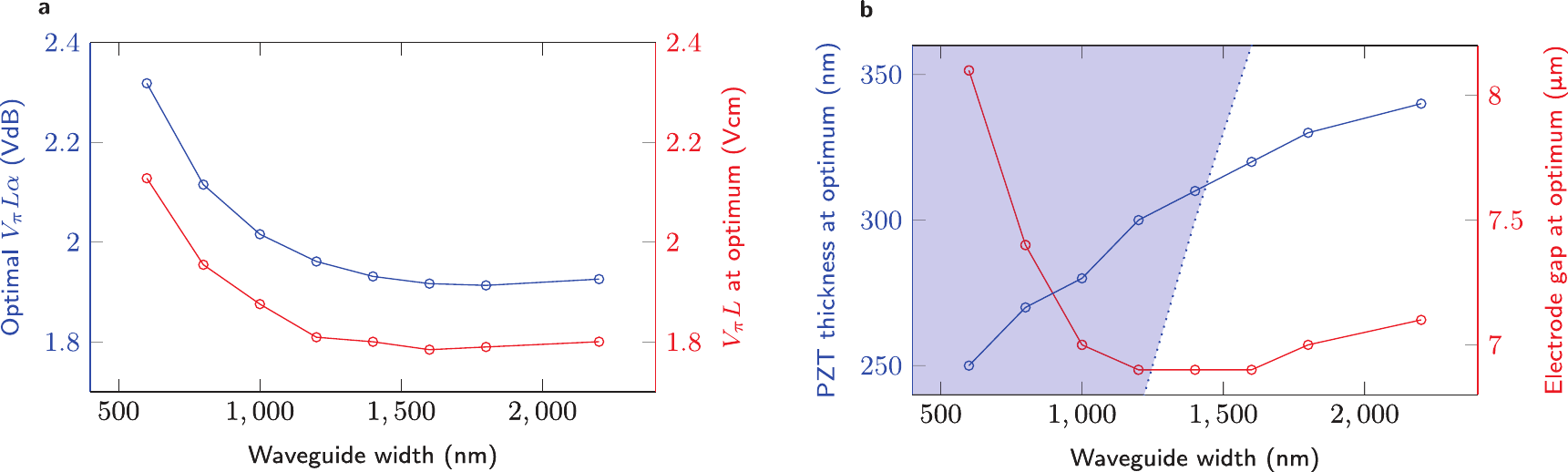}
  \caption{\sffamily\footnotesize \textbf{Numerical optimization of the a PZT-on-SiN modulator as a function of waveguide width.}  Optimization of the waveguide loss times the half-wave voltage-length product $V_\pi L \alpha$ of a PZT-covered SiN waveguide modulator as a function of waveguide width. \textbf{a}, For each waveguide width $V_\pi L \alpha$ is minimized as function of both electrode spacing and PZT thickness (blue line). The red line plots the calculated $V_\pi L$ at this optimum. \textbf{b}, Electrode spacing (blue) and PZT thickness (red) at the optimum. The light blue area shows the waveguide width/PZT thickness combinations for which the waveguide only supports a single TE mode. Wavelength, waveguide height and intermediate layer thickness are respectively 1550 nm, 300 nm and 20 nm. The intrinsic waveguide loss (in the absence of electrodes) was taken to be 1 dB/cm, the effective electro-optic coefficient 67 pm/V.}\label{fig:figS5}
\end{figure*}

In the simulations in Fig. \ref{fig:fig4}, a sweep of the electrode spacing and PZT thickness was performed, since these can be easily tailored in post-processing. This was done for a fixed waveguide width of 1.2 \textmugreek m. The waveguide width can however also be designed, an optimization is given here. At each width, a sweep of $V_\pi L \alpha$ as a function of PZT thickness and electrode gap of the kind described in the main text and shown in Fig. \ref{fig:fig4} was performed. Fig. \ref{fig:figS5}a shows the optimal (smallest) value  $\mathrm{min}(V_\pi L \alpha)$ and the $V_\pi L$ at that optimum. Fig. \ref{fig:figS5}b shows the PZT thickness and electrode spacing of this optimum. The light blue area shows the waveguide width/PZT thickness combinations for which the waveguide only supports a single TE mode. In the main text, a width of 1.2 \textmugreek m was chosen in order to minimize $\mathrm{min}(V_\pi L \alpha)$ whilst still having single-mode behavior at the optimal point.
\end{document}